\documentclass[twocolumn,prb,amsfonts,amsmath,amssymb,floatfix]{revtex4} 
\usepackage{color}
\usepackage{soul}
\usepackage{hhline}	
\usepackage{mathrsfs}
\usepackage{graphicx}
\usepackage{dcolumn}
\usepackage{bm}
\usepackage{multirow}
\usepackage{booktabs}
\usepackage{afterpage}
\usepackage{amsmath}
\usepackage{braket}
\usepackage{array}

\begin{document}

\title{Temperature- and doping-dependent roles of valleys in thermoelectric performance of SnSe: a first-principles study}

\author{Hitoshi Mori}
\author{Hidetomo Usui}
\author{Masayuki Ochi}
\author{Kazuhiko Kuroki}
\affiliation{Department of Physics, Osaka University, Machikaneyama-cho, Toyonaka, Osaka 560-0043, Japan}

\date{\today}
\begin{abstract}
We theoretically investigate how each orbital and valley play a role for high thermoelectric performance of SnSe.
In the hole-doped regime, two kinds of valence band valleys contribute to its transport properties: one is the valley near the U-Z line, mainly consisting of the Se-$p_z$ orbitals, and the other is the one along the $\Gamma$-Y line, mainly consisting of the Se-$p_y$ orbitals.
Whereas the former valley plays a major role in determining the transport properties at room temperature, the latter one also offers comparable contribution and so the band structure exhibits multi-valley character by increasing the temperature.
In the electron-doped regime, the conduction band valley around the $\Gamma$ point solely contributes to the thermoelectric performance, where the quasi-one-dimensional electronic structure along the $a$-axis is crucial. This study provides an important knowledge for the thermoelectric properties of SnSe, and will be useful for future search of high-performance thermoelectric materials.
\end{abstract}
\pacs{}

\maketitle

\section{Introduction}

Thermoelectric effects have gathered much attention because of its technological importance, e.g., for electric power generation.
Many researchers have attempted to make efficient use of thermal energy, which is ubiquitous but is not sufficiently exploited, as a promising energy resource.
The efficiency for thermoelectric conversion is characterized by the dimensionless figure of merit $ZT$:
\begin{align}
ZT &= \frac{\sigma S^2}{\kappa}T = \frac{\mathrm{PF}}{\kappa}T, \\
\mathrm{PF} &= \sigma S^2,
\end{align}
where $\sigma$, $S$, $\kappa=\kappa_{\mathrm{el}}+\kappa_{\mathrm{ph}}$, $\mathrm{PF}$, and $T$ are the electrical conductivity, Seebeck coefficient, thermal conductivity, which is often separated into the electronic and phonon contribution, powerfactor, and temperature, respectively.
For example, ${\rm Bi}_{2}{\rm Te}_{3}$~\cite{BiTe-1,BiTe-2,BiTe-3}, Pb$X$($X$: chalcogens such as Te and Se)~\cite{PbTe-1,PbTe-3,PbTe-4,PbSe-1}, and ${\rm Co}_{}{\rm Sb}_{3}$~\cite{CoSb3-1,CoSb3-2,CoSb3-3,CoSb3-4,CoSb3-5}, exhibit a $ZT$ value larger than one, and are now used for industrial applications.
However, a wider range of applications requires a larger value of $ZT$, which is a central objective in studies of thermoelectric materials.

To achieve a high $ZT$, thermoelectric materials should simultaneously possess a high powerfactor and a low thermal conductivity.
To realize this situation, optimization of the carrier concentration is crucial. This is because the electrical conductivity, the Seebeck coefficient, and the inverse of the thermal conductivity have different maxima with respect to the carrier concentration. For example, whereas the electrical conductivity becomes larger by an increase of the carrier concentration, the thermal conductivity also increases owing to the Wiedemann-Franz low. In addition, the Seebeck coefficient usually decreases by going away from the band edge because of a decreased difference of the group velocities between the electron and hole carriers around the chemical potential.

In 2014, it was reported that $p$-type SnSe exhibits a very large $ZT \sim 2.6$ at $T\sim 920$ K~\cite{Nature (London)li}.
SnSe is a layered material with the orthorhombic (space group: $Pnma$) and cubic (space group: $Cmcm$) lattices, below and above the structural transition temperature $T_c=807$ K, respectively.
Its high $ZT$ owes to a very low thermal conductivity $\kappa \sim 0.23{\rm -}0.24\, {\rm W}^{}{\rm m}^{-1}{\rm K}^{-1}$ and a moderate powerfactor ${\rm PF}\sim10\, \mu {\rm W}^{}{\rm cm}^{-1}{\rm K}^{-2}$.
This finding had a great impact on the field of thermoelectrics. 
From the experimental side, several studies for optimizing the carrier concentration in single crystals or polycrystalline samples have been reported both for the hole~\cite{p-type-poly-SnSe-Ag-doped,p-type-poly-SnSe,p-type-single-SnSe-Na-doped-1,p-type-single-SnSe-Na-doped-2,p-type-single-SnSe-Ge-doped} and electron~\cite{n-type-poly-SnSe-S-doped,n-type-poly-SnSe-BiCl3-doped,n-type-single-SnSe-Bi-doped} carriers. In particular, a recent study reported that $ZT \sim 2.2$ was achieved in $n$-type SnSe single crystal~\cite{n-type-single-SnSe-Bi-doped}.
Many theoretical studies have also been conducted on the electronic and phonon properties of SnSe~\cite{SnSe-phonon-calc,SnSe-espresso-calc,SnSe-wien-calc,SnSe-vasp-calc,kktr,rg,SnSe-prl-calc,rlg}.
The maximum value of $ZT$ with the optimum carrier concentration is an important issue also for the theoretical studies, and some of these studies suggested that the powerfactor of $n$-type SnSe can exceed that for $p$-type SnSe.
The theoretical studies also revealed that the band structure with several valleys, which consist of the Sn-$5p$ and Se-$4p$ orbitals, affects its thermoelectric property in a rather non-trivial way, and the role of each orbital and valley has not been well understood.

\begin{figure*}
 \begin{center}
\centering
  \includegraphics[width=16cm,clip]{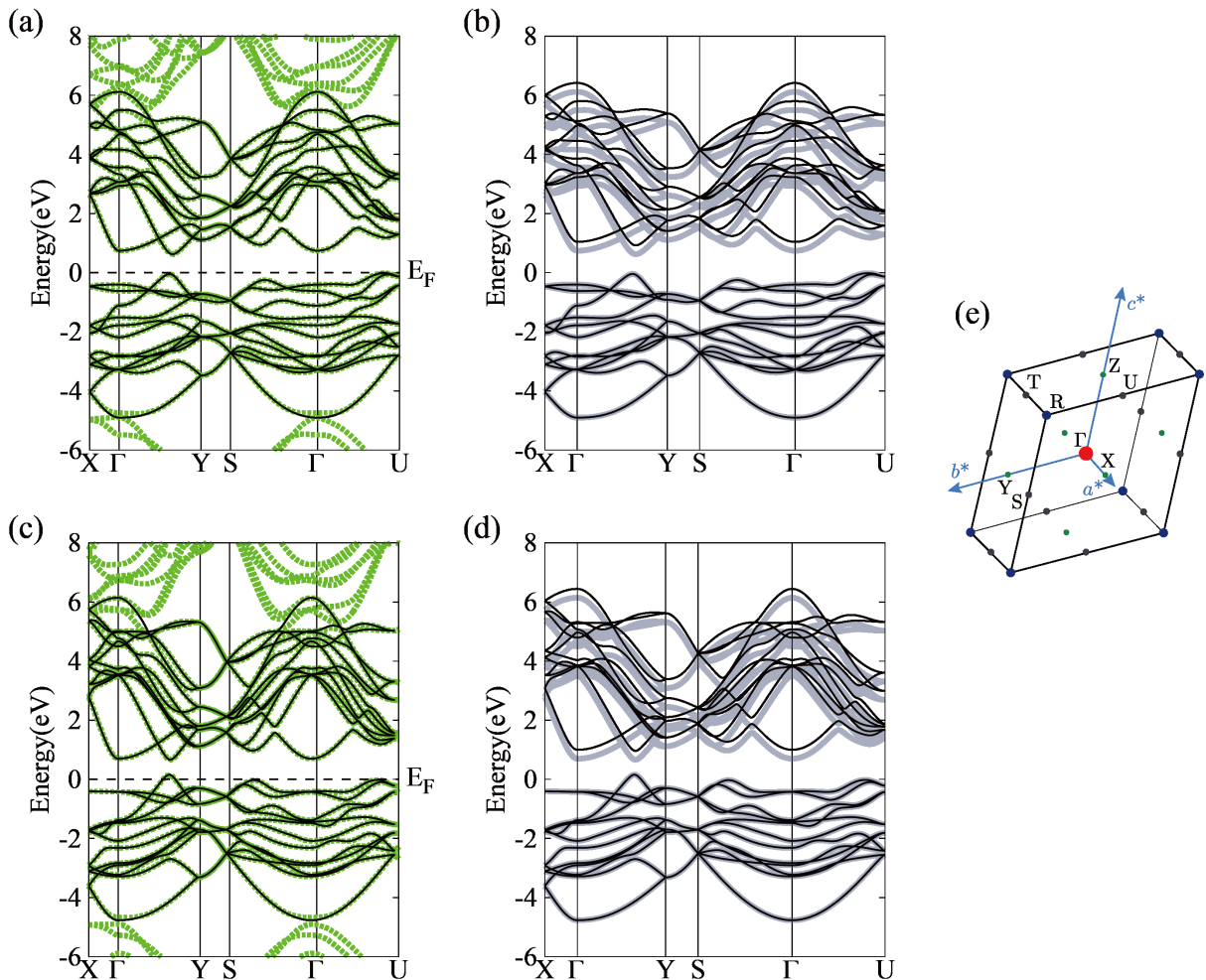}
  \caption{(a) Electronic band structures obtained with first-principles calculation (green broken lines) and model calculation (black solid lines) for the 295 K crystal structure. Black thin lines in panel (b) is the same as that in panel (a), and gray thick lines denote that with an upward shift of the conduction bands by 0.3 eV, which is used for calculations of transport properties in this study. Those for the 790 K crystal structure are presented in panels (c) and (d). The Brillouin zone for the $Pnma$ phase is shown in panel (e).}
   \label{fig:1}
 \end{center}
\end{figure*}

In this study, we investigate the thermoelectric properties, particularly the powerfactor, of SnSe with a detailed analysis on the role of each orbital and valley by means of the first-principles calculations.
To enable the comparison between the theoretical and experimental values of transport quantities, we focus on the $Pnma$ phase in this study.
In the hole-doped regime, the powerfactor along the $b$-axis is the highest among all directions.
In this regime, the powerfactor is enhanced by the valley near the U-Z line (cf. Fig.~\ref{fig:1}(e)) at the room temperature.
On the other hand, at a higher temperature around the structural phase transition, the valley near the $\Gamma$-Y line also enhances the electrical conductivity, which means that the transport properties are governed by the multi-valley band structure.
In the electron-doped regime, the powerfactor along the $a$-axis is the highest because strongly anisotropic electron carriers, which have a much lower effective mass along the $a$-axis than the other directions, reside in the conduction band valley around the $\Gamma$ point.
Contribution from the conduction bands around the $\Gamma$ point becomes more dominant by increasing the temperature.

This paper is organized as follows. Section~\ref{sec:model} presents the way to construct an effective model employed in our analysis.
Boltzmann transport theory used in our study is described in Sec.~\ref{sec:boltz}.
We verify the consistency between the experimental results and our theoretical analysis in Sec.~\ref{sec:exp}.
Theoretical investigation for the dependency of the transport properties on the carrier concentration at 750 K is presented in Sec.~\ref{sec:dope}. Section~\ref{sec:orb} describes the contribution of each orbital to the transport coefficients, and additional analysis of the relationship between the powerfactor and the band structure is presented in Sec.~\ref{sec:origin}. Section~\ref{sec:sum} summarizes this study.

\section{Construction of the effective model\label{sec:model}}

First, we performed the first-principles band structure calculation of SnSe in the $Pnma$ phase using the WIEN2k package~\cite{wien}.
Experimental crystal structures at 295 K ($a=11.501$ \AA, $b=4.153$ \AA, and $c=4.445$ \AA) and 790 K ($a=11.621$ \AA, $b=4.334$ \AA, and $c=4.282$ \AA) were extracted from Ref.~[\onlinecite{chtt}]. Here, we investigated the crystal structures at two different temperatures, in order to discuss how the temperature change affects the thermoelectric properties.
The Perdew-Burke-Ernzerhof parametrization of the generalized gradient approximation~\cite{pbegga} was used without including the spin-orbit coupling because we verified that the spin-orbit coupling has only a small effect on the band structures. We set $RK_{\mathrm{max}}=7$ and used a $4\times 11\times 10$ $k$-mesh.
Figure~\ref{fig:1}(a)(c) presents the calculated band structures for the crystal structures at 295 K and 790 K, respectively.
These results are consistent with the previous theoretical studies~\cite{kktr,rg,rlg}.

Next, we extracted the maximally localized Wannier functions~\cite{wannier01,wannier02,wannier1,wannier2} for the Sn-5$p$ and Se-4$p$ orbitals from these first-principles band structures. Using these Wannier functions, we constructed a 24-band tight-binding models for these orbitals, the band structures obtained from which are shown in Fig.~\ref{fig:1}(a)(c).
Hereafter, we analyze the transport properties using these effective models and the Boltzmann transport theory described in the next section.
Note that we shift the conduction bands upward by 0.3 eV in order to make the band gap consistent with the experimental value~\cite{Nature (London)li} as shown in Fig.~\ref{fig:1}(b)(d).
To simulate the carrier doping, we adopt the rigid band approximation.

\section{Boltzmann transport theory\label{sec:boltz}}

In the Boltzmann transport theory, tensor quantities
\begin{align}
{\boldsymbol \sigma}&=e^2{\bf K}_0  , \\
{\bf S}&=-\frac{1}{eT}{\bf K}_0^{-1}{\bf K}_1  , \label{eq:seebeck}\\
{\boldsymbol \kappa}_{\rm el}&=\frac{1}{T}\left[{\bf K}_2-{\bf K}_1{\bf K}_0^{-1}{\bf K}_1 \right] ,
\end{align}
are represented with the transport coefficients ${\bf K}_{\nu}$:
\begin{widetext}
\begin{align}
{\bf K}_\nu= \tau\sum_n\sum_{\bm{k}} \bm{v}_n(\bm{k})\otimes\bm{v}_n(\bm{k})\left[-\frac{\partial f_0(\epsilon_n(\bm{k}),T )}{\partial \epsilon} \right](\epsilon_n(\bm{k})-\mu(T))^\nu ,
\end{align}
\end{widetext}
where $e$ $(>0)$, $f_0$, $n$, and $\bm{k}$ are the elementary charge, the Fermi-Dirac distribution function at equilibrium, the band and the $k$-point indices, respectively.
The group velocity $\bm{v}_n$ and the energy level $\epsilon_n$ for the $n$-th band can be obtained by diagonalizing the tight-binding Hamiltonian constructed in the way described in the previous section.
In this study, the relaxation time $\tau$ was assumed to be constant by neglecting its dependence on the crystal momentum, band index, and temperature.
We set $\tau=3.0\times10^{-15}\,$sec, which was determined by fitting the calculated powerfactor to the experimental one along the $b$-axis at 750 K (see Fig.~\ref{fig:2}(c)).

\section{Results and Discussion}

\subsection{Consistency with the experimental results\label{sec:exp}}

\begin{figure*}
 \begin{center}
  \includegraphics[width=16cm]{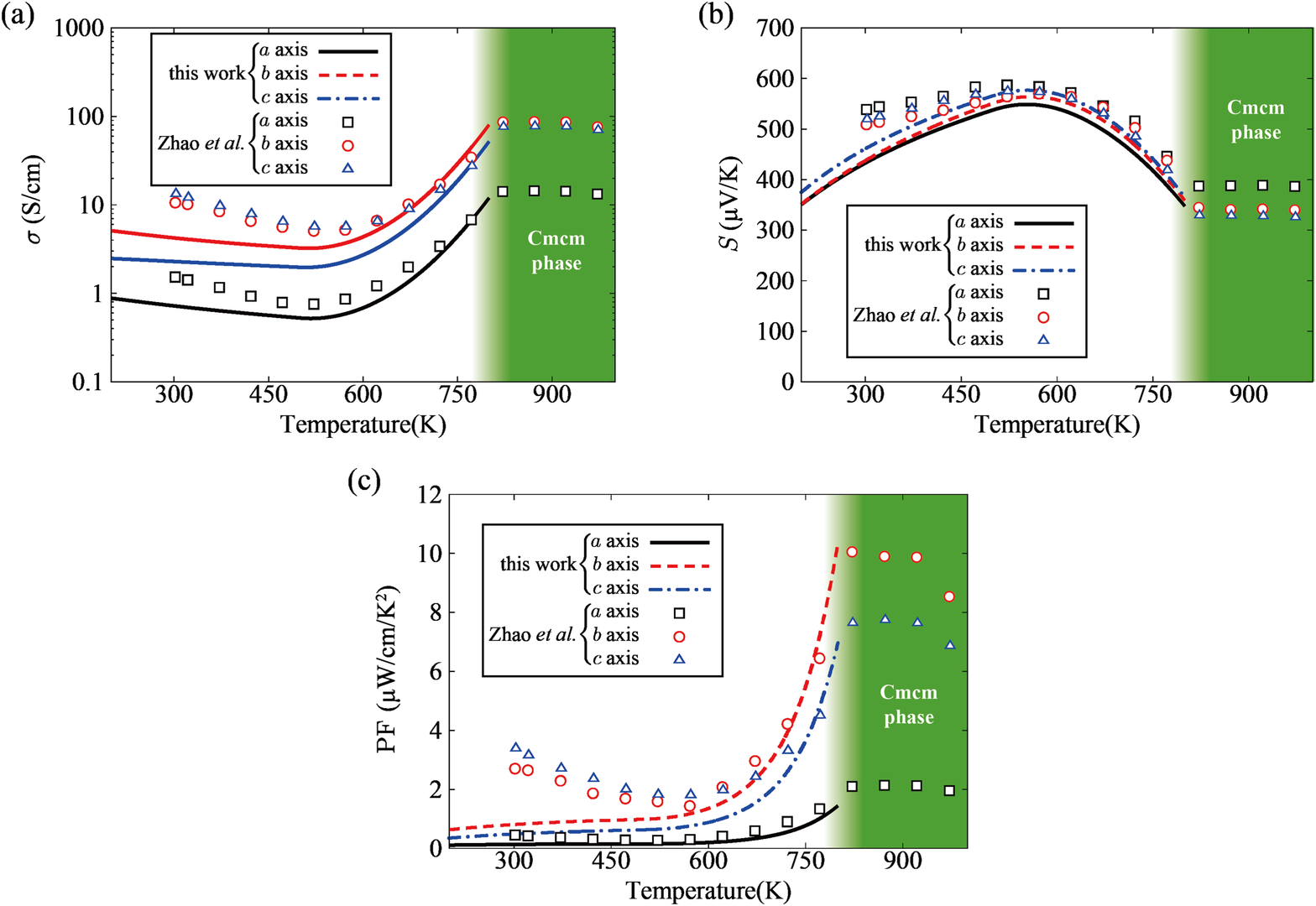}
  \caption{Temperature dependence of (a) the electrical conductivity $\sigma$, (b) the Seebeck coefficient $S$, and (c) the powerfactor PF with the hole-carrier concentration presented in Fig.~\ref{fig:3}. Experimental results taken from Ref.~[\onlinecite{Nature (London)li}] are also shown all the panels. }
   \label{fig:2}
 \end{center}
\end{figure*}

\begin{figure}
 \begin{center}
  \includegraphics[width=8.2cm]{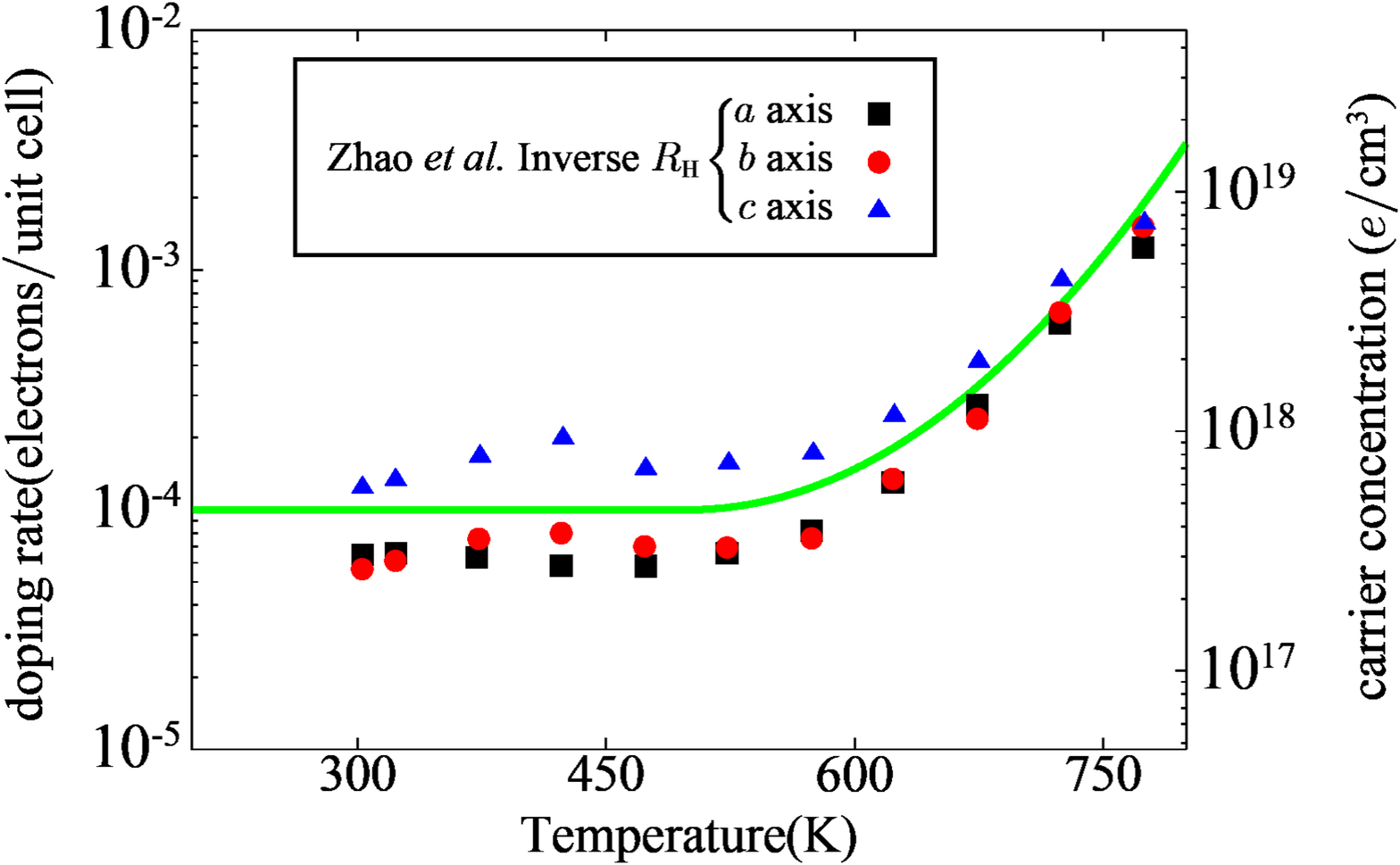}
  \caption{Hole-carrier concentration used for plots shown in Fig.~\ref{fig:2}, together with experimental values of the inverse Hall coefficients taken from Ref.~[\onlinecite{Nature (London)li}].}
   \label{fig:3}
 \end{center}
\end{figure}

First, we checked whether our theoretical analysis well reproduces the experimental transport properties.
In this subsection, we employed the band structure obtained from the crystal structure at 790 K.
Because the experimental observation on the Hall coefficient suggests that the carrier concentration becomes higher by increasing the temperature~\cite{Nature (London)li}, we assumed its temperature dependence as follows:
\begin{align}
n (T)=\left\{
\begin{array}{ll}
n_0 & (T<T_0)\\
n_0\exp\left[\left(\frac{T-T_0}{T_1}\right)^2\right] & (T\geq T_0)
\end{array}\right.
\end{align}
where $n_0=0.0001$ $e/{\rm u.c.}$, $T_0=500$ K, and $T_1=160$ K.
As shown in Fig.~\ref{fig:3}, this function well represents the experimental variation of the inverse Hall coefficient in temperature. Using this carrier concentration, we calculated the temperature dependence of the electrical conductivity $\sigma$, Seebeck coefficient $S$, and powerfactor PF as presented in Fig.~\ref{fig:2}.
Whereas the electrical conductivity, and thereby the powerfactor, are underestimated at low temperature in our simulation owing to the constant relaxation-time approximation, these quantities at high temperature and the Seebeck coefficients in all the temperature range below the structural phase transition temperature well reproduce the experimental ones.
Because we determined the relaxation time so that the calculated and experimental powerfactors along the $b$-axis are consistent at 750 K, it can be naturally understood that such relaxation time is underestimated in low temperature where the scattering events become less frequent. We also note that the relaxation time cancels out for the Seebeck coefficient, Eq.~(\ref{eq:seebeck}), under the constant relaxation-time approximation. By looking into the temperature variation of these transport quantities, we can see that the anisotropy of the powerfactor mainly comes from that of the electrical conductivity.

\subsection{Doping dependence\label{sec:dope}}

\begin{figure*}
 \begin{center}
  \includegraphics[width=16cm]{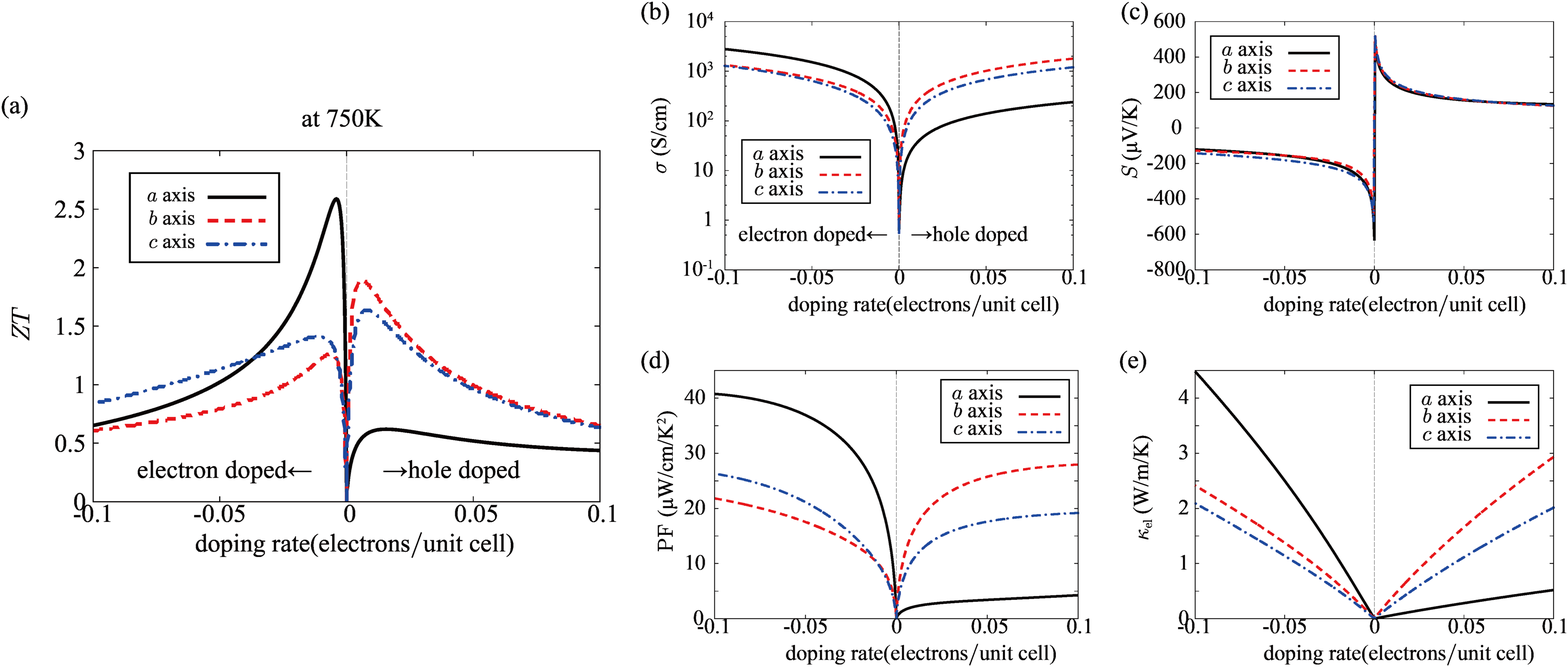}
  \caption{Doping dependence of (a) the dimensionless figure of merit $ZT$, (b) the electrical conductivity $\sigma$, (c) the Seebeck coefficient $S$, (d) the powerfactor PF, and (e) the electrical thermal conductivity $\kappa_{\mathrm{el}}$ at 750 K.}
   \label{fig:4}
 \end{center}
\end{figure*}

Using the calculated band structure obtained from the crystal structure at 790 K, we simulated the doping dependence of several transport quantities at 750 K as shown in Fig.~\ref{fig:4}.
Here, we assumed the lattice thermal conductivity $\kappa_{\mathrm{ph}}$ to be constant against the carrier doping rate. The value of $\kappa_{\mathrm{ph}}$ was determined by subtracting the calculated electrical thermal conductivity $\kappa_{\mathrm{el}}$ from the total thermal conductivity extracted from the experiment~\cite{Nature (London)li} at 750 K: $\kappa_{a\text{-axis}}^{\rm exp}=0.22$ ${\rm W}^{}{\rm m}^{-1}{\rm K}^{-1}$, $\kappa_{b\text{-axis}}^{\rm exp}=0.33$ ${\rm W}^{}{\rm m}^{-1}{\rm K}^{-1}$, and $\kappa_{c\text{-axis}}^{\rm exp}=0.29$ ${\rm W}^{}{\rm m}^{-1}{\rm K}^{-1}$.

In the hole-doped regime, the electrical conductivity and the electrical thermal conductivity along the $b$-axis are the largest and those along the $a$-axis are the smallest among all directions.
On the other hand, in the electron-doped regime, these quantities become the largest along the $a$-axis.
The Seebeck coefficients are nearly isotropic in both regimes.
As a result, the dimensionless figure of merit $ZT$ becomes the largest along the $b$- and $a$-axes in the hole- and electron-doped regimes, respectively.
In particular, the $ZT$ value along the $a$-axis in the electron-doped regime reaches $\sim 2.5$.
These tendencies, i.e., the anisotropy of transport quantities and the fact that the highest $ZT$ is achieved along the $a$-axis in the electron-doped regime, are consistent with the previous theoretical studies~\cite{kktr,rg}.

\subsection{Orbital decomposition of transport coefficients\label{sec:orb}}

To obtain a deeper insight into the transport properties with the non-trivial anisotropy as seen in the previous subsection, we performed the orbital decomposition of the transport coefficients as follows:
\begin{widetext}
\begin{align}
{\bf K}^{(\alpha)}_\nu= \tau\sum_n\sum_{\bm{k}} |C_{\alpha n}(\bm{k})|^2\bm{v}_n(\bm{k})\otimes\bm{v}_n(\bm{k})\left[-\frac{\partial f_0(\epsilon_n(\bm{k}),T )}{\partial \epsilon} \right](\epsilon_n(\bm{k})-\mu(T))^\nu ,
\end{align}
\end{widetext}
where the coefficients $C_{\alpha n}(\bm{k})$ for each orbital $\alpha$ is defined as
\begin{align}
\ket{\psi_{n\bm{k}}}=\sum_{\alpha}C_{\alpha n}(\bm{k})\sum_{\bm{R}}e^{i\bm{k}\cdot\bm{R}}\ket{\bm{R}_{\alpha}},
\end{align}
for the Bloch state in the $n$-th band at the crystal momentum ${\bm k}$, which is represented as linear combination of the Wannier basis $\ket{\bm{R}_{\alpha}}$ indexed with the lattice vector ${\bm R}$, and satisfies the eigenvalue equation for the tight-binding Hamiltonian $\mathcal{H}_{\rm eff}$:
\begin{align}
\mathcal{H}_{\rm eff}\ket{\psi_{n\bm{k}}}=\epsilon_{n}(\bm{k}) \ket{\psi_{n\bm{k}}}.
\end{align}
Here, ${\bf K}^{(\alpha)}_\nu$ represents the contribution from the $\alpha$ orbital to ${\bf K}_\nu$ and obviously satisfies
\begin{equation}
\sum_{\alpha} {\bf K}^{(\alpha)}_\nu = {\bf K}_\nu.
\end{equation}

\begin{figure*}
 \begin{center}
  \includegraphics[width=17cm]{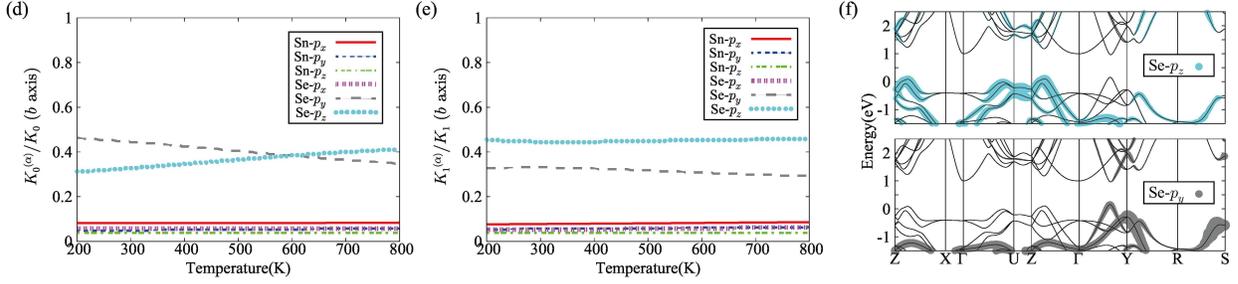}
  \caption{(a), (b), (d), (e) Temperature dependence of the ratio of orbital contribution to the transport coefficients $K^{(\alpha)}_{\nu}/K_{\nu}$ ($\nu = 0, 1$) along the $b$-axis with a hole-doping rate $n=0.01$ $e/{\rm u.c.}$ (c), (f) Band structures with a colored weight of the Se-$p_z$ orbital: $|C_{(\text{Se-}p_z) n}(\bm{k})|^2$ and Se-$p_y$ orbital: $|C_{(\text{Se-}p_y) n}(\bm{k})|^2$. Panels (a)-(c) employ the 295 K crystal structure, and other panels for the 790 K crystal structure.}
   \label{fig:5}
 \end{center}
\end{figure*}

As we have seen in the previous subsection, the $ZT$ value is the highest along the $b$-axis in the hole-doped regime.
To investigate its origin in more detail, we calculated the ratio of the orbital contribution to the transport coefficients $K^{(\alpha)}_{\nu}/K_{\nu}$ ($\nu = 0, 1$) along the $b$-axis with a hole-doping rate $n=0.01$ $e/{\rm u.c.}$ as shown in Fig.~\ref{fig:5}.
To see the temperature effect, we performed calculations for two experimental crystal structures at 295 K and 790 K.
Whereas the Se-$p_z$ orbital mainly contributes to the transport coefficients in Fig.~\ref{fig:5}(a)(b) using the 295 K crystal structure, the Se-$p_y$ orbital also has a comparable contribution in Fig.~\ref{fig:5}(d)(e) using the 790 K crystal structure. Contributions from these two orbitals have larger temperature dependence for $K_0$ than that for $K_1$.
Because $K_0$ and $K_1$ are affected by a different size of the energy region around the chemical potential, this behavior suggests that the energies of the relevant band structures relative to the chemical potential are crucial in determining the transport properties of SnSe in the hole-doped regime.
By looking into the band structures presented in Fig.~\ref{fig:5}(c)(f), we can see that the Se-$p_z$ and Se-$p_y$ orbitals have a large weight near the U-Z line and midway between the $\Gamma$ and Y points, respectively.

\begin{figure*}
 \begin{center}
  \includegraphics[width=17cm]{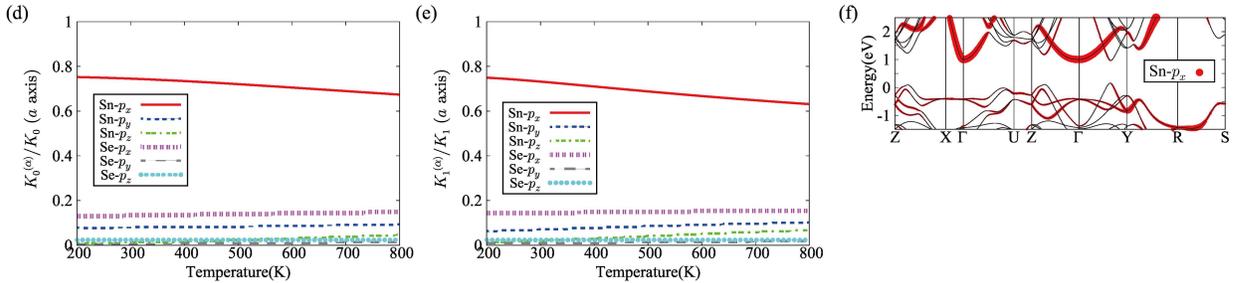}
  \caption{(a), (b), (d), (e) Temperature dependence of the ratio of orbital contribution to the transport coefficients $K^{(\alpha)}_{\nu}/K_{\nu}$ ($\nu = 0, 1$) along the $a$-axis with an electron-doping rate $n=-0.01$ $e/{\rm u.c.}$ (c), (f) Band structures with a colored weight of the Sn-$p_x$ orbital: $|C_{(\text{Sn-}p_x) n}(\bm{k})|^2$. Panels (a)-(c) employ the 295 K crystal structure, and other panels for the 790 K crystal structure.}
   \label{fig:6}
 \end{center}
\end{figure*}

In the electron-doped regime, we performed a similar calculation along the $a$-axis, where the $ZT$ value reaches the maximum, with an electron-doping rate $n=-0.01(e/{\rm u.c.})$ as shown in Fig.~\ref{fig:6}.
We can see that the Sn-$p_x$ orbital always play a dominant role for the transport coefficients. This tendency is enhanced in the 790 K crystal structure (Fig.~\ref{fig:6}(d)(e)) compared with the 295 K one (Fig.~\ref{fig:6}(a)(b)) .
Figure~\ref{fig:6}(c)(f) shows that the transport properties are governed by the band dispersion around the $\Gamma$ point, where a large weight of the Sn-$p_x$ orbital can be found.

\subsection{Origin of the large powerfactor\label{sec:origin}}

To clarify the complex role of the two different orbitals, Se-$p_y$ and $p_z$, for the transport properties and their relevance with the band structures in the hole-doped regime, we decomposed the transport coefficients in terms of the $k$-space region as follows:
\begin{widetext}
\begin{align}
K_\nu&= \tau\sum_n\sum_{\bm{k}\in \rm{B.Z.}} v^2(\bm{k})\left[-\frac{\partial f_0(\epsilon_n(\bm{k}),T )}{\partial \epsilon} \right](\epsilon_n(\bm{k})-\mu(T))^\nu \label{KBZs} \\ 
&=K_\nu^{({\rm Z})}+K_\nu^{(\Gamma)}, \notag
\end{align}
where
\begin{eqnarray}
\begin{cases}
\displaystyle K_\nu^{({\rm Z})}= \tau\sum_n\sum_{\bm{k},\,\, |k_z|>\pi/(2c)} v^2(\bm{k})\left[-\frac{\partial f_0(\epsilon_n(\bm{k}),T )}{\partial \epsilon} \right](\epsilon_n(\bm{k})-\mu(T))^\nu & \\
\displaystyle K_\nu^{(\Gamma)}= \tau\sum_n\sum_{\bm{k},\,\, |k_z|\leq\pi/(2c)} v^2(\bm{k})\left[-\frac{\partial f_0(\epsilon_n(\bm{k}),T )}{\partial \epsilon} \right](\epsilon_n(\bm{k})-\mu(T))^\nu & \end{cases}.\label{KBZe}
\end{eqnarray}
\end{widetext}
Owing to the band structures near the valence band top as presented in Fig.~\ref{fig:5}(c)(f), $K_\nu^{({\rm Z})}$ and $K_\nu^{(\Gamma)}$ can be regarded as the transport coefficients with contributions from the regions near the U-Z and $\Gamma$-Y lines, respectively.
In our simulation, $\mu(T)$ in Eqs.~(\ref{KBZs}), (\ref{KBZe}) were determined in common by considering occupations in all band dispersion.
By using these decomposed transport coefficients, we calculated the powerfactors defined as
\begin{eqnarray}
{\rm PF}=\frac{1}{T^2}\frac{K_1^2}{K_0},\,\, \begin{cases}\vspace{5pt}\displaystyle {\rm PF}^{({\rm Z})}=\frac{1}{T^2}\frac{\left(K_1^{({\rm Z})}\right)^2}{K_0^{({\rm Z})}} & \\ \displaystyle {\rm PF}^{(\Gamma)}=\frac{1}{T^2}\frac{\left(K_1^{(\Gamma)}\right)^2}{K_0^{(\Gamma)}}& \end{cases},
\end{eqnarray}
where we should note that ${\rm PF}\neq{\rm PF}^{({\rm Z})}+{\rm PF}^{(\Gamma)}$.

\begin{figure}
 \begin{center}
  \includegraphics[width=8.2cm]{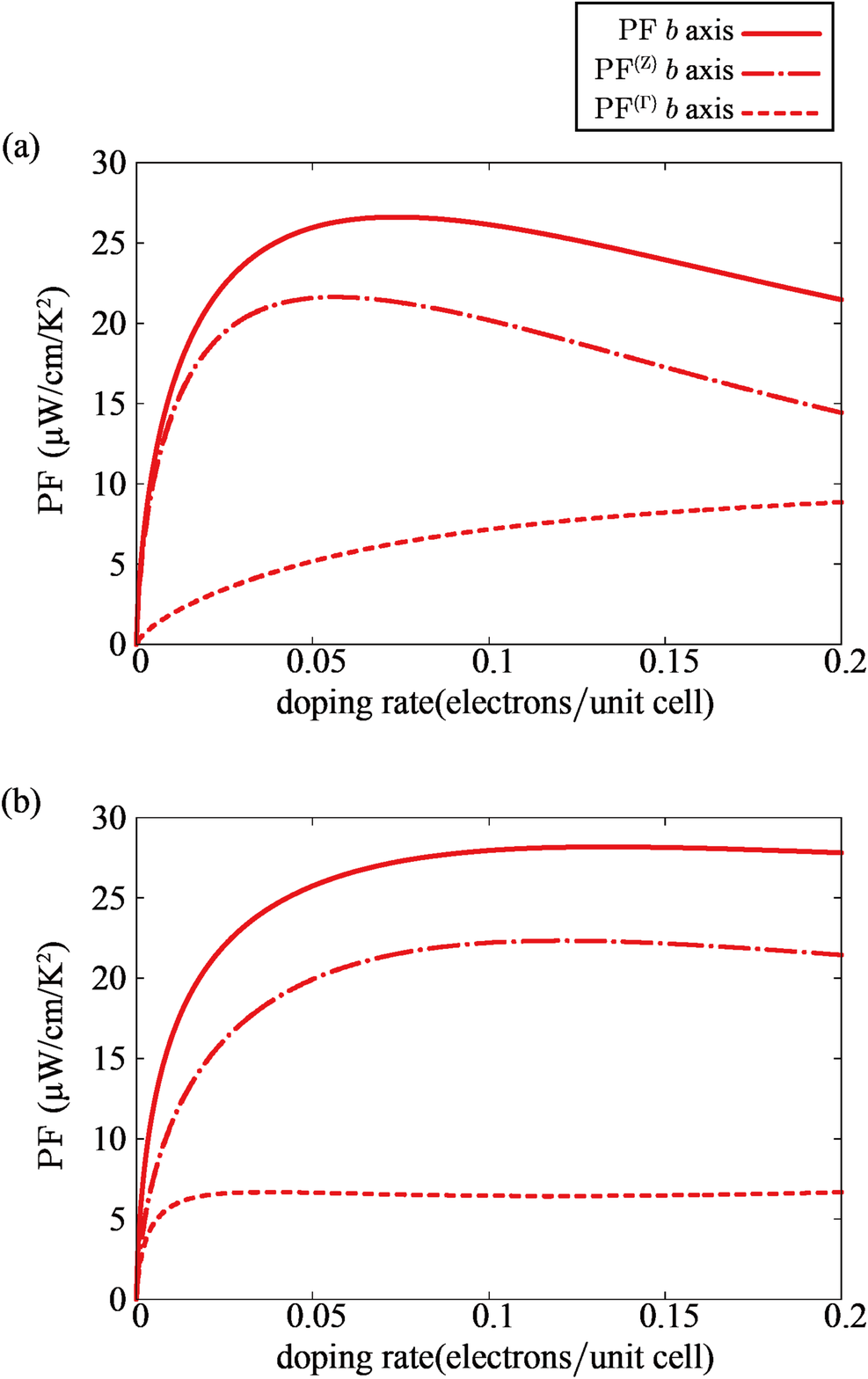}
  \caption{Doping dependence of the powerfactors along the $b$-axis at 750 K using the (a) 295 K and (b) 790 K crystal structures.}
   \label{fig:7}
 \end{center}
\end{figure}

Figure~\ref{fig:7} presents the calculated powerfactors along the $b$-axis using the two crystal structures.
When the hole-carrier concentration is small, the total powerfactors exhibit only a small difference between two structures.
For example, the total powerfactor PF$=15.9$ $\mu {\rm W}^{}{\rm cm}^{-1}{\rm K}^{-2}$ for the 295 K crystal structure and 16.3 $\mu {\rm W}^{}{\rm cm}^{-1}{\rm K}^{-2}$ for the 790 K crystal structure, with the hole-carrier doping rate $n=0.01$ $e/{\rm u.c.}$ On the other hand, ${\rm PF}^{({\rm Z})}=14.2$ $\mu {\rm W}^{}{\rm cm}^{-1}{\rm K}^{-2}$ for the 295 K crystal structure but $11.0$ $\mu {\rm W}^{}{\rm cm}^{-1}{\rm K}^{-2}$ for the 790 K crystal structure.
By such a difference, in the small hole-carrier concentration regime, the band structure for the 790 K crystal structure exhibits the multi-valley character~\cite{multi} for its transport properties whereas that for the 295 K crystal structure does not (i.e. the band dispersion along the U-Z line has a dominant role).
This comes from a change of the relative energy levels between the valence band valleys around the U-Z line and that along the $\Gamma$-Y line.
Note that the maximum values of ${\rm PF}^{({\rm Z})}$ with respect to the doping rate are similar between the two crystal structures: 21.6 and 22.3 $\mu {\rm W}^{}{\rm cm}^{-1}{\rm K}^{-2}$ for the 295 K and 790 K crystal structures, respectively; that is, the valleys around the U-Z line for both the crystal structures have apparently different shapes but almost the same potential for the powerfactor. This means that the smaller ${\rm PF}^{({\rm Z})}$ for the 790 K structure than that for the 295 K structure with $n=0.01$ $e/{\rm u.c.}$ is caused not by deteriorating the shape of the valleys for thermoelectric performance but by a change of the distance between the chemical potential and the valleys around the U-Z line.
On the other hand, the valley along the $\Gamma$-Y line becomes closer to the chemical potential in the 790 K crystal structure even with a low carrier concentration, which compensates a decrease of ${\rm PF}^{({\rm Z})}$ and provides a similar value of the total powerfactor PF for the two crystal structures.
Note that, by seeing the whole behavior presented in Figs.~\ref{fig:5} and \ref{fig:7}, the band dispersion near the U-Z line seems more favorable and has a larger contribution for increasing the powerfactor compared with that along the $\Gamma$-Y line.
K. Kutorasinski {\it et al.}~\cite{kktr} also pointed out that the band structure along the U-Z line has a pudding-mold-like shape, which is favorable for the thermoelectric performance~\cite{krk}.
\begin{figure}
 \begin{center}
  \includegraphics[width=8.2cm]{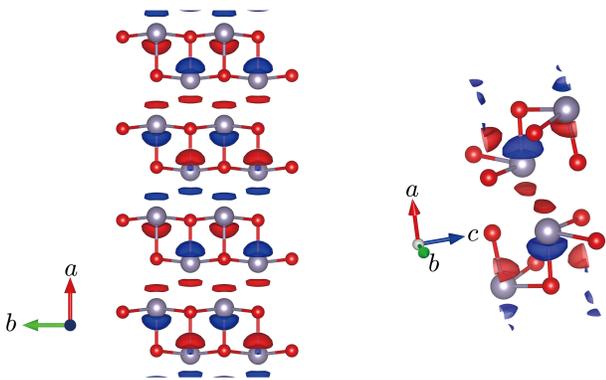}
  \caption{Bloch state on the lowest conduction band at the $\Gamma$ point depicted using the VESTA software~\cite{VESTA}.}
   \label{fig:8}
 \end{center}
\end{figure}

In the electron-doped regime, the situation is completely different.
Thermoelectric properties in that regime are solely determined by the conduction band valley around the $\Gamma$ point, especially by its quite anisotropic electronic structure.
Indeed, we calculated the effective mass at the $\Gamma$ point and found its strong anisotropy: $m_{xx}/m_0=0.069$, $m_{yy}/m_0=0.84$, and $m_{zz}/m_0=2.11$, where $m_0$ is the mass of free electrons and calculation was performed around the $\Gamma$ point, $k_i\in [0,0.1\pi]$ ($i=x,y,z$).
Such anisotropic effective mass was also calculated by K. Kutorasinski {\it et al.}~\cite{kktr} and R. Guo {\it et al.}~\cite{rg}
We can easily see that the group velocity along the $\Gamma$-X line is much larger than that along the $\Gamma$-Y and $\Gamma$-Z lines for the lowest conduction band as shown in Fig.~\ref{fig:6}(e)-(f).
Because a one-dimensional band structure is shown to be more favorable for high thermoelectric performance than two- and three-dimensional band structures~\cite{usuione}, we can say that the quasi-one-dimensional band structure here is the origin of the large powerfactor in the electron-doped regime.
The corresponding Bloch state on the lowest conduction band at the $\Gamma$ point obtained from our first-principles calculation is shown in Fig.~\ref{fig:8}.
As is consistent with our orbital analysis in the previous subsection, the Bloch state mainly consists of the Sn-$p_x$ orbitals and has a quasi-one-dimensional character, which is enabled by an inter-layer transfer along the $a$-axis between Sn atoms.

\section{Conclusion\label{sec:sum}}

We have investigated the non-trivial roles of the valleys in the band structure for thermoelectric performance of SnSe.
In the hole-doped regime, the valence band valley near the U-Z line, mainly consisting of the Se-$p_z$ orbitals, and the one along the $\Gamma$-Y line, mainly consisting of the Se-$p_y$ orbitals, contribute to the large powerfactor along the $b$-axis in the manner dependent on the temperature.
Whereas the powerfactor is mainly determined from the former valley in the crystal structure around the room temperature, the latter valley gets closer to the chemical potential and increases its contribution to the transport properties in the crystal structure near the structural phase transition.
Even with such multi-valley character, the valley near the U-Z line is more important for the high powerfactor.
In the electron-doped regime, the conduction band valley around the $\Gamma$ point solely contributes to the thermoelectric performance, where the quasi-one-dimensional electronic structure along the $a$-axis is realized. This tendency is enhanced by increasing the temperature through a change of the crystal structure.
This study deepens our understanding of the thermoelectric properties of SnSe, and will be useful for further improving its performance.

\acknowledgments
This study was supported by JSPS KAKENHI (Grant No. JP26610101) and  JST CREST (Grant No. JPMJCR16Q6). We appreciate fruiteful discussion with N.  Hanasaki, H. Sakai, H. Murakawa, K. Katayama, T. Sakamoto, H. Li, and T. Nishimura. Valuable comments from D. Ogura are also gratefully acknowledged.


\begin{thebibliography}{999}
\bibitem{multi}We note that we refer to the relation of the valence band valleys along the $\Gamma$-Y and U-Z lines as `multi-valley' and discuss it in this paper. On the other hand, each valley along the $\Gamma$-Y and U-Z lines also has trivial multi-valley character owing to the space inversion symmetry.
\end{thebibliography}
\end{document}